**Phase Stability in Ferroelectric Bismuth Titanate: A First Principles Study**

*Anurag Shrinagar[1], Ashish Garg[1], and Rajendra Prasad[2]*

[1]Department of Materials and Metallurgical Engineering

[2]Department of Physics

Indian Institute of Technology Kanpur

Kanpur 208016, India

**Synopsis:**

'This work is a report on the investigation of room temperature structural stability of a technologically important ferroelectric oxide bismuth titanate where, using first principle calculations, it is shown that monoclinic structure with space group *B1a1* is the ground state structure.'.

**Abstract:**

Experimental data on the structure of ferroelectric oxide Bismuth Titanate suggests two different kinds of structures *i.e.* orthorhombic and monoclinic. We have performed density functional theory (DFT) based first principle calculations to determine the ground state structure of bismuth titanate, based on experimentally observed monoclinic and orthorhombic phases of $Bi_4Ti_3O_{12}$. Both of these phases are optimized to zero pressure and lattice parameters were determined as *a* = 5.4370 Å, *b* = 5.4260 Å, *c* = 32.6833 Å and Z = 4 for structure with space group *B2cb* and *a* = 5.4289 Å, *b* = 5.4077 Å, *c* = 32.8762 Å, *β* = 90.08° and Z = 4 for the structure with space group *B1a1* . Static and dynamic calculations show that the monoclinic structure with the space group *B1a1* is the ground state structure. It is noted that small difference in the energies of both structures could be a factor behind experimental observation of either of the structure.

## 1. Introduction:

Bismuth layered compounds such as $SrBi_2Ta_2O_9$ and $Bi_4Ti_3O_{12}$ belong to the Aurivillius family of phases [1,2,3] and are denoted by a general formula $(Bi_2O_2)^{2+}(A_{n-1}B_nO_{3n+1})^{2-}$ where $n$ represents the number of perovskitic $(A_{n-1}B_nO_{3n+1})^{2-}$ layers which are alternatingly stacked with fluorite like $(Bi_2O_2)^{2+}$ layers along the $c$-axis of the unit cell.

Bismuth titanate, $Bi_4Ti_3O_{12}$, is one such material with $n = 3$ and has attracted tremendous amount of attention in the past decade, primarily due to its potential for nonvolatile memory applications as its thin films showed good ferroelectric properties and more importantly, lower processing temperatures over its predecessor $SrBi_2Ta_2O_9$ [4,5]. However, the structure of $Bi_4Ti_3O_{12}$ has been intriguing because experimental studies report either monoclinic or orthorhombic structured phases. First, in 1949, Aurivillius, based on X-ray diffraction results, conceived its structure as orthorhombic with space group *Fmmm* and lattice parameters $a = 5.410$ Å, $b = 5.448$ Å and $c = 32.8$ Å [1]. Later, in 1971, Dorrian *et al* [6] showed that although the X-ray diffraction data supported an orthorhombic structure with space group *B2cb* and lattice parameters $a = 5.448(2)$, $b = 5.411(2)$, $c = 32.83(1)$ Å, physical properties indicated the monoclinic structure. The authors also observed a small but finite polarization along $c$-axis which is an unlikely event because in the *B2cb* structure, **b** glide has its mirror plane perpendicular to the $c$-axis so projections of all the polarization vectors along $c$-axis get cancelled.

In 1990, Rae *et al.* [7] conducted structure refinement of bismuth titanate using electron diffraction data obtained on single crystal bismuth titanate and reported the structure to be monoclinic with space group *B1a1* and lattice parameters $a = 5.450(1)$, $b = 5.4059(6)$, $c = 32.832(3)$ Å and $\beta = 90°$. In *B1a1*, the absence of **b** glide plane explained the existence of finite remnant polarization along the $c$-axis. Subsequent refinement of neutron diffraction data by Hervoches and Lightfoot [8] suggests the structure to be orthorhombic with space group *B2cb* and lattice parameters $a = 5.4444(1)$, $b = 5.4086(1)$ and $c = 32.8425(6)$ Å. More recent experimental observations have reported the monocline structured phase with space group *B1a1* [9,10]. Above findings show that the experimentally determined room temperature structures of bismuth titanate show subtle differences and are dependent upon the history of the sample. These disparities emphasize upon the need for a theoretical study to investigate the structure of bismuth titanate.

Previous structural optimization of the $Bi_4Ti_3O_{12}$ was performed by Noguchi et al. [11] on I4/mmm tetragonal structure but the tetragonal phase of $Bi_4Ti_3O_{12}$ is not a room temperature phase and is stable only above its Curie temperature of 948 K [12]. To the best of our knowledge, no structural optimization has been performed on the room temperature structures of $Bi_4Ti_3O_{12}$ *i.e.* either monoclinic or orthorhombic. Such a structure optimization will also be important in resolving some of the disagreements observed in the lanthanide doped bismuth titanate thin films regarding polarization in *c*-axis oriented epitaxial thin films [13,14,15] and one of the reasons that could lead to observed disparities is substrate induced strain. In this context, a study on structure prediction will further enable us to theoretically investigate the structural changes in this material that occur upon doping and application of external strain, especially in thin film form.

In this paper, we present a first principle study using density functional theory on the experimentally observed room temperature phases of $Bi_4Ti_3O_{12}$. In our study, we have used the crystallographic data of three experimental studies on $Bi_4Ti_3O_{12}$, performed by Dorrian et al. [6], Rae et al. [7] and Hervoches et al. [8]. We conducted our calculations of the pressure and total energy on these experimental structures. We have also calculated the optimized structures of orthorhombic and monoclinic phases.

## 2. Computational details:

The calculations for optimization are performed in the framework of first principles density functional theory [16]. Vienna ab-initio simulation package (VASP) [17,18,19] is used for structural optimization in the present study using projector augmented wave method (PAW) [20]. The Kohn-Sham equations [21,22] are solved using the exchange correlation function of Perdew and Wang [23] for generalized gradient approximation (GGA) scheme. A plane wave energy cut-off of 400 eV is used. We used Monkhorst-Pack [24] sampling using the 4×4×4 mesh. Conjugate Gradient [25] algorithm is used for the structural optimization. In relaxation, Gaussian smearing width of 0.1 eV is used for the determination of partial occupancies. Tetrahedron method with Blöchl corrections [26] is used to determine the total energy.

Static calculations are performed on the experimental structures keeping their volume, cell shape and ionic positions fixed. First of all, in dynamic calculations, ions are relaxed into their instantaneous ground state keeping the volume and shape of the unit cell fixed. Later, cell shape is relaxed while keeping its volume and position of ions fixed to get the optimized structure. This optimized structure is relaxed repeatedly through same process and relaxation procedure continues until we get the stable ionic positions up to accuracy of $10^{-4}$ in fractional coordinates. Symmetry is kept fixed in fractional coordinates to the accuracy of $10^{-5}$ during the relaxation process to restore the correct charge density and forces. Forces between the ions are relaxed below 0.005 eV/Å since below 0.01 eV/Å, ions are supposed to be energetically stable. Volume of the unit cell is changed manually by adjusting the cell constant in the input to get the total pressure equal to zero as well as zero pressure along *x*, *y* and *z* axes of Cartesian coordinate system.

### 3. Results and Discussion:

The idealized structure of bismuth titanate having space group *Fmmm*, as shown in Figure 1, was first observed by Aurivillius [1]. The structure consists of $Bi_2O_2^{2+}$ layers alternating with perovskite structured $Bi_2Ti_3O_{10}^{2-}$ layers. However, displacement of A cations ($Bi^{3+}$) along with corresponding cooperative tilting and distortion of $TiO_6$ octahedra causes deviation from the ideal structure giving rise to the observed ferroelectricity in this compound [6]. Subsequent experimental studies carried out using X-ray, neutron and electron diffraction methods showed that room temperature structure of bismuth titanate has space group either *B2cb* [6,8] or *B1a1* [7]. We began our studies with these three experimental structures as suggested by Dorrian *et al.* [6] [structure A thereafter], Hervoches and Lightfoot [8] [structure B thereafter] and Rae *et al.* [7] [structure C thereafter]. We first calculated the stability of these structures on the basis of energy and the pressure by conducting static calculations, keeping the parameters of all unit cells and their corresponding ionic positions fixed and the results are shown in table 1.

First we will analyze and compare the structures A and B having similar space groups. Table 1 shows that structure A exhibits higher values of both energy and pressure as compared to structure B, an indication of comparatively high compressive stress on structure A. We also compared the lattice parameters of these two structures and the values are shown in Table 2 and it can be seen from the table that cell parameters of the unit cell of structure A are nearly same to the cell parameters of unit cell of structure B. To further analyze this, we analyzed the ionic positions and the forces on each ion, as shown in Table 3. The table shows that the forces on every ion in the both structures are quite high and both of these structures are not stable. Particularly higher values of forces are observed on Ti(1), Ti(2), O(1),

O(5) and O(6) ions in the structure A in comparison to structure B. These ions are also depicted in the structure shown in figure 2. The figure shows that O(5) and O(6) are the corner ions of the octahedra in the upper and lower layers with Ti(2) located at the center of these octahedra. O(1) ion is located at the corners of octahedra in central layer and Ti(1) is located at the center of these octahedra. Comparatively higher values of forces on these ions in structure A arises because of significant differences in their positions as compared to structure B. These results indicate that the coordinates determined by Hervoches et al.[8] for structure B are more precise than those determined by Dorrian et al.[6] for structure A. In addition, large energy differences between the structures A and B, 8.88 eV per 2 f.u., again supports the rather precise nature of the ionic positions determined by Hervoches et al.[8]. On the basis of these observations, we did not consider structure A any further and chose structure B with space group *B2cb* for further optimization along with structure C with space group *B1a1*.

To reduce the time of calculations, we preferred to conduct our calculations on the primitive unit cells of both the structures. Since both of these structures are B-centered and contain 2 lattice points, primitive unit cell of these structures were chosen with axes $a' = (a-c)/2$, $b' = b$, $c' = (c+a)/2$. Change in unit cell also modifies the fractional coordinates from the fraction of B-centered unit cell's axes to the fraction of primitive unit cell's axes. Both, primitive and parent unit cells follow the condition that the arrangement of ions remains same in the Cartesian coordinate system as shown in the equation (1):

$$x.\vec{a} + y.\vec{b} + z.\vec{c} = x'.\vec{a}' + y'.\vec{b}' + z'.\vec{c}' \qquad [1]$$

where $(x, y, z)$ are the fractional coordinates of an ion with respect to $(a, b, c)$ axes of the B-centered non-primitive unit cell and $(x', y', z')$ are fractional coordinates with respect to $(a', b', c')$ axes of the primitive unit cell. The $a'$, $b'$ and $c'$ axes are written in terms of $a$, $b$ and $c$ in the following manner:

$$\vec{a}' = (\vec{a} - \vec{c})/2 \qquad [2]$$

$$\vec{b}' = \vec{b} \qquad [3]$$

$$\vec{c}' = (\vec{a} + \vec{c})/2 \qquad [4]$$

Regardless of the values of $\vec{a}$, $\vec{b}$ and $\vec{c}$ in the Cartesian coordinate system, the solution of above four equations is always $x' = x-z$, $y' = y$ and $z' = x+z$. So after calculating this data, our optimization is conducted finally on the primitive unit cell of both these structures which essentially contains total of 38 ions. The primitive unit cell of a structure with space group *B1a1* is equivalent to its original Bravais lattice with space group *P1n1* in which the equivalent positions for the space group

*B1a1*, (x, y, z) and **a**:(x+1/2, -y+1/2, z), change to (x, y, z) and **n**: (x+1/2, -y+1/2, z+1/2) as shown in figure 3.

Relaxation of structures B and C yields the optimized structures which we have denoted as structures 'D' (space group *B2cb*) and 'E' (space group *B1a1*) respectively. Energy and cell parameters of the structure B, C, D and E are reported in table 1 and table 2 respectively. Fractional coordinates of structure B and C are shown in table 3 and that of structure D and E are shown in table 4. In the final representation of structure D, coordinate axes have been rotated by 180° about *c*-axis and the origin is shifted by **a/2** and **b/2** along +*a*-axis and +*b*-axis respectively. This change of coordinate axes changes the (x, y, z) coordinates of all the ions to (-*x*+1/2, -*y*+1/2, *z*). Coordinate axes are changed to obtain the same format of the fractional coordinates in both structures D and E. It is noticed from table 1 that after relaxation the energy of the structure D (space group *B2cb*) decreases by 1.87 eV per 2 f.u. *i.e.* from structure B to structure D. This decrease in the energy creates a few subtle changes in the structure D from its parent structure B as seen in the tables 1-4. Table 2 shows that the *c*-axis length of structure D decreases by 0.16 Å but rest of the cell parameters of structure D do not change appreciably. Comparison of tables 3 and 4 shows that in structure D, ionic coordinates of O(6) change substantially: O(6) is displaced along +*a*-axis by 0.79 Å and along *b*-axis by 0.56 Å on both sides of the **c**-glide plane in the opposite directions. This change is also shown by arrows in the corresponding structures in figure 4(a) and 4(b), projected on $(0\bar{1}0)$ plane. Rest of the ions do not show any significant change. In structure B, Ti(2) ions are displaced along +*a*-axis and Ti(1) ions are displaced along -*a*-axis from the center of their oxygen octahedra. But after optimization, in structure D, the observed shift of O(6) along +*a*-axis makes the Ti(2) ions also displace along –*a*-axis from center of their oxygen octahedra and therefore optimized *B2cb* unit cell shows the same polarization direction along *a*-axis as that in structure C and E.

After optimization of structure C to structure E with space group *B1a1*, the energy decreases by 0.44 eV per 2 f.u.. The relatively smaller change in the energy leads to very close resemblance between the structures C and E in terms of lattice parameters (table 2) and atomic positions (table 4). The optimized structure E is also represented in figure 4(c) projected on $(0\bar{1}0)$.

Now we can compare the optimized structure D (space group *B2cb*) with structure E (space group *B1a1*). On the basis of lower energy as shown in table 1, it can be stated that structure E is more stable than structure D and is the ground state structure. However, by studying the coordinates of ions in structures D and E as seen in table 4, we observed a close resemblance between both the structures. Major differences are observed only in the coordinates of O(1) and O(1)′ ions. The $x$ and $y$ coordinates of O(1) of structures D and E differ by 0.18 Å and 0.22 Å and that of O(1)′ differ by 0.25 Å and 0.25 Å respectively. For other ions, the differences in coordinates are comparatively very low. These observations suggest that these differences in the ionic positions are because of presence of **2**-fold axis and **b**-glide plane in the *B2cb* structure and when these symmetries are allowed to relax, the coordinates re-adjust themselves fractionally to yield a more stable structure E (space group *B1a1*). In the parent structure (space group *Fmmm*), the O and Ti ions located at $x=1/4$ and $x=1/2$ which are shifted along $+a$-axis in both *B2cb* and *B1a1* structures (B & C or D & E) (also refer to figure 4).

Results of calculations on bond lengths and bond angles of structure D and structure E are shown in tables 5 and 6. Bond angles and bond lengths of both structures D and E are quite close to each other since the ionic coordinates and cell parameters of both of these structures are very similar as seen in table 4 and table 2 respectively. An analysis of the bond lengths and bond angles of both structures show that there is a significant difference in the two bond lengths of Ti(2)—O(6), Ti(2)'—O(6)', Ti(2)—O(5), Ti(2)'—O(5)', Ti(1)—O(1) and Ti(1)—O(1)'. Therefore all the $Ti^{4+}$ ions are shifted along $-a$-axis from the center of their oxygen octahedra. The displacement of $Ti^{4+}$ ions from the center of their oxygen octahedra for the central layer octahedra is much larger as compared to those in the upper and lower layer octahedra *e.g.* in structure D (space group *B2cb*), the bond length difference in two bonds of Ti(1)—O(1) is 0.163 Å while the bond length differences of Ti(2)—O(5) and Ti(2)—O(6) are 0.122 and 0.096 Å respectively. Hence the bond length difference of Ti(1)—O(1) bonds is more than the bond length differences of Ti(2)—O(5) and Ti(2)—O(6) bonds, giving rise to larger shift of the Ti(1) along $-a$ axis than Ti(2). Same can also be observed in structure E (space group *B1a1*). The shift of Ti along $-a$-direction is the main reason for higher values of polarization along $a$-axis as observed in many experimental studies. In the structure E, Ti(2)—O(4) & Ti(2)′—O(4)′, Ti(2)—O(3) & Ti(2)′—O(3)′ and Ti(1)—O(3) & Ti(1)—O(3)′ bond lengths in the upper, central and lower octahedra are not equal. But in structure D, because of the presence of **b**-glide plane, these bond lengths are equal. In this structure, the magnitude of the shifts of Ti in upper octahedron and lower octahedron is same but in opposite directions along the $c$-axis and in central octahedron, the Ti is placed at the center along $c$-axis, thus causing the polarization along $c$-axis to be zero in *B2cb* structure D. On the other hand, in structure E with *B1a1* space group, the bond lengths in the upper, central and lower octahedra are not equal. Therefore in *B1a1* phase, finite value of polarization is expected along the *c*-axis.

In the end, however, comparison of energies and close resemblance in the atomic positions of structure D and structure E suggest that it is likely that room temperature structure may be any of these two. The situation may further be complicated by the volatility of Bi leading to Bi-nonstoichiometry and any oxygen nonstoichiometry which may change the energetics and the ultimate stability of the structure.

**4. Conclusions:**

We have made an analysis on the structural stability of ferroelectric oxide bismuth titanate by the first principles density functional calculations using the experimental data[6,7,8]. First the precision of structures from different experimental studies was compared and analyzed via static calculations made on the basis of energy and pressure. The results suggested that the the structure with *B2cb* space group (B) predicted by Hervoches *et al.*[8] was more precise than the similar structure (A), shown previously by Dorrian *et al.*[6], latter exhibiting higher values of forces on its ions and higher total energy. However, structure with *B1a1* space group (C) predicted by Rae *et al.*[7] appeared to be more precise than the structure B with space group *B2cb*. Further, dynamic calculations were carried on structures B and C to yield the optimized structures D and E respectively. We found the structure with space group *B1a1* (E) to be the ground state structure with minimum energy. How, it must be noted, that although, structure E with *B1a1* space group has lower energy than structure D with *B2cb* space group, the energy difference between both of these two structures is very small, also confirmed by the close resemblance between the unit cell parameters and fractional coordinates of both structures. Study of bond lengths and bond angles of *B1a1* structure E reveals that because of the absence of **b**-glide, polarization along the *c*-axis would also be observed in this structure which is found absolute zero in *B2cb* structure D. We observed that all Ti ions are shifted towards -*a*-axis in each octahedron indicating the presence of a finite polarization along *a*-axis. We also note that the energetically close nature of both the structures, *B1a1* and *B2cb*, could be reason why experimental studies show either of these structures at room temperature.

**Acknowledgements:**

We have our deep gratitude towards Dr. Arjit Sen and Dr. Shailesh Shukla, who have been very helpful in carrying out this whole research.

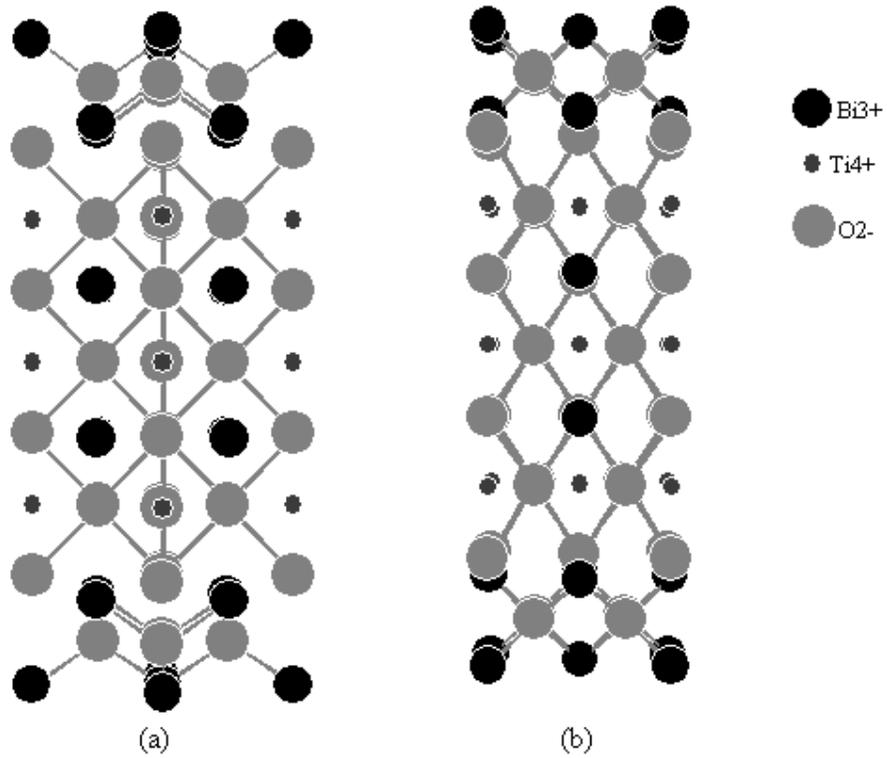

**Figure 1: (a) A perspective drawing of *Fmmm* parent phase as viewed along (110) projection (b) as viewed along (100) or (010) projection plane. Only half of the ions between c=¼ and c=¾ are shown.**

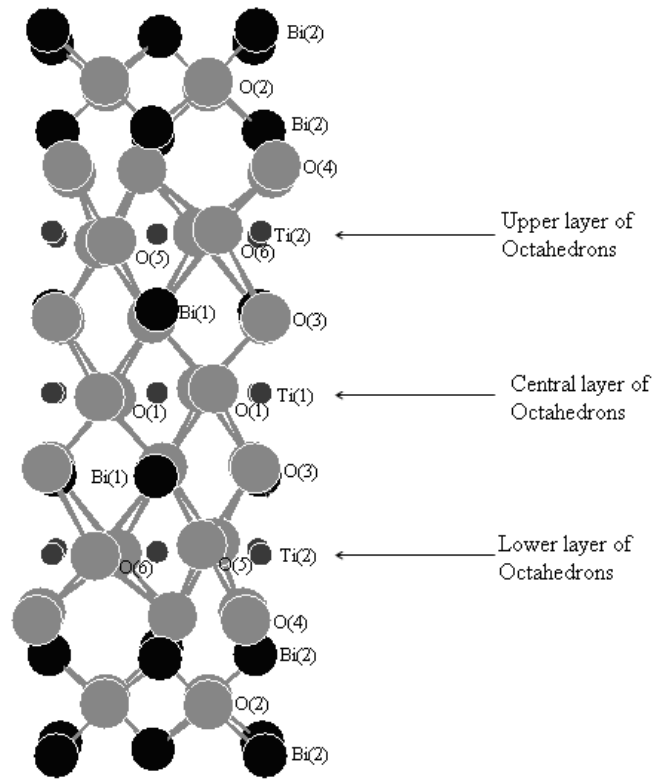

**Figure 2:** *B2cb* **structure as observed by Dorrian *et al.* projected on (100) plane. Ions are shown between 1/4c and 3/4c only.**

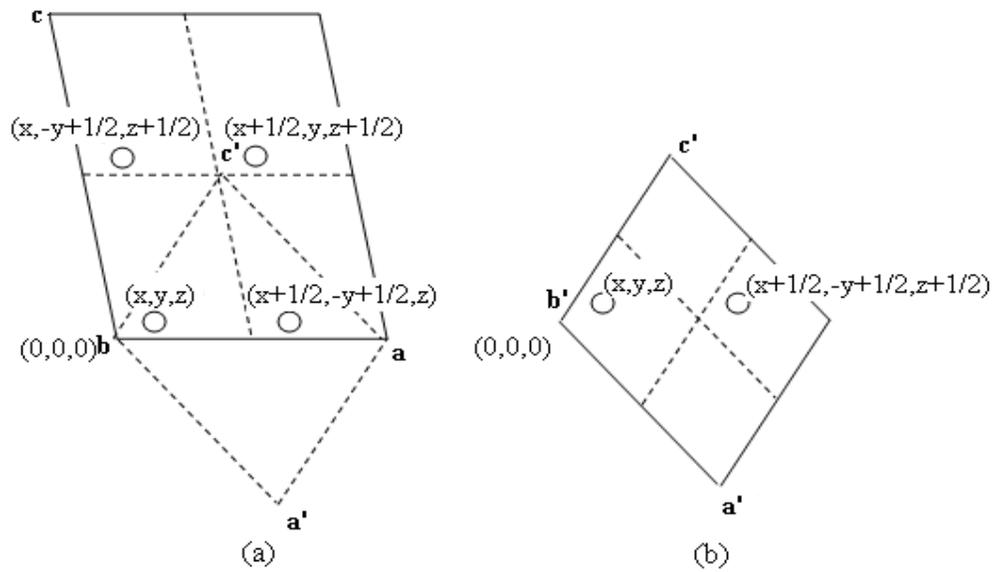

**Figure 3:** (a) Solid lines show B-centered unit cell and circles show the general equivalent points generated by *B1a1* operation. Dashed lines show the simple monoclinic unit cell with a', b' and c' axes (b) general equivalent points of *P1n1* space group in unit cell with a', b' and c' axes. n glide is located at *y*=1/4.

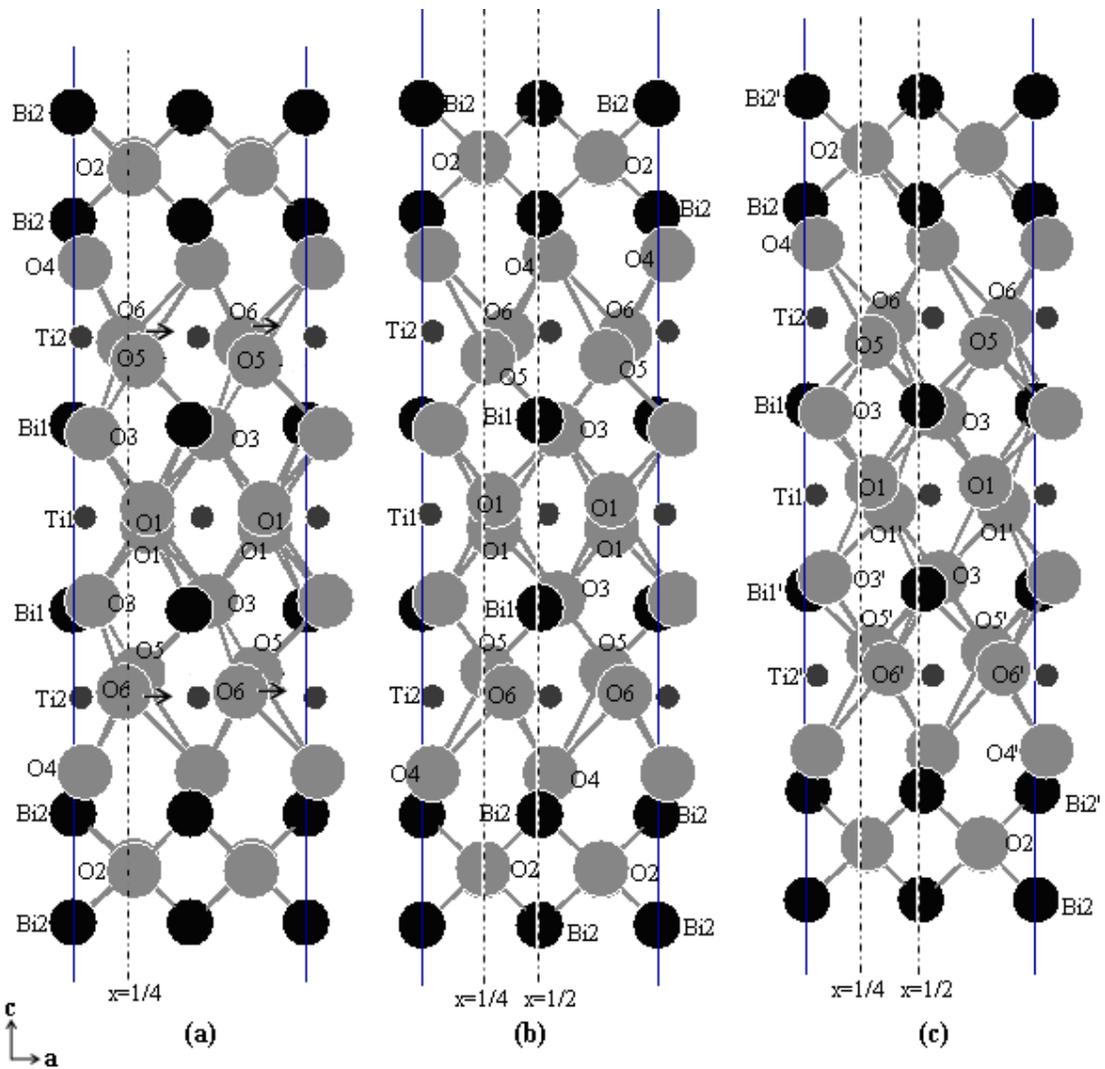

**Figure 4:** (a) Schematic representation of structure B. In this structure arrows show the displacement of O(6) ions observed after optimization. (b) Schematic representation of structure D (c) Schematic representation of structure E. In all the structures, ions are shown only between 1/4c and 3/4c and projection of unit cells is taken on $(0\bar{1}0)$ plane. For *B2cb* structure ( )' are shown by ( ) only since they are equivalent (x, -y, -z) points.

**Table 1: Results of static and dynamic calculations conducted on experimental structures. Static calculation were performed on structure A, B and C whereas dynamic calculations yielded structure D and E.**

| Identification | Space Group | Total Energy Per 2 f. u. (eV)* | External Pressure on the lattice (kB) | Reference |
|---|---|---|---|---|
| Structure A | B2cb | -266.50 | 60.5 | Dorrian et al. [6] |
| Structure B | B2cb | -275.38 | 16.0 | Hervoches et al. [8] |
| Structure C | B1a1 | -276.97 | 20.0 | Rae et al. [7] |
| Structure D | B2cb | -277.25 | 0 | Optimized B structure |
| Structure E | B1a1 | -277.31 | 0 | Optimized C structure |

*f.u.: formula units

**Table 2: Lattice parameters of experimental structures and optimized structures**

| Identification | a(Å) | b (Å) | c(Å) | β | space group |
|---|---|---|---|---|---|
| Structure A | 5.448(2) | 5.411(2) | 32.83(1) | - | B2cb (orthorhombic) |
| Structure B | 5.4444(1) | 5.4086(1) | 32.8425(6) | - | B2cb (orthorhombic) |
| Structure C | 5.450(1) | 5.4059(6) | 32.832(3) | 90° | B1a1 (monoclinic) |
| Structure D | 5.4370 | 5.4260 | 32.6833 | - | B2cb (orthorhombic) |
| Structure E | 5.4289 | 5.4077 | 32.8762 | 90.08° | B1a1 (monoclinic) |

**Table 3:** Fractional coordinates and total forces (eV/Å) on the experimental structures under study. For structure C (*B1a1* space group) only initial coordinates (x, y, z) are reported but for structure A and B (*B2cb* space group) equivalent (x, -y, -z) coordinates are also reported.

|       | Structure A | | | | Structure B | | | | Structure C | | | |
|-------|---|---|---|---|---|---|---|---|---|---|---|---|
| **Ion** | **x** | **y** | **z** | **F** | **x** | **y** | **z** | **F** | **x** | **y** | **z** | **F** |
| Bi(1)  | 0 | 0.5022(2) | 0.5668(0) | 1.99 | 0 | 0.5018(7) | 0.56639(8) | 0.44 | 0.0030(1) | 0.5023(1) | 0.5673(1) | 0.48 |
| Bi(1)'* | 0 | -0.5022(2) | -0.5668(0) | 1.99 | 0 | 0.4982(7) | 0.43361(8) | 0.44 | 0.0013(1) | 0.4977(1) | 0.4336(1) | 0.27 |
| Bi(2)  | -0.0009(3) | 0.4801(1) | 0.7114(0) | 0.9 | 0.001(1) | 0.4861(9) | 0.71127(8) | 0.68 | -0.0021(1) | 0.4793(1) | 0.7113(0) | 1.07 |
| Bi(2)' | -0.0009(3) | -0.4801(1) | -0.7114(0) | 0.9 | 0.001(1) | 0.5139(9) | 0.28873(8) | 0.68 | 0.0021(1) | 0.5185(1) | 0.2887(0) | 0.25 |
| Ti(1)  | 0.0452(8) | 0 | 0.5 | 4.47 | 0.052(2) | 0 | 0.5 | 0.78 | 0.0446(2) | -0.0013(6) | 0.5007(2) | 0.58 |
| Ti(2)  | 0.0533(6) | 0.0001(1) | 0.6286(1) | 6.61 | 0.037(2) | -0.004(2) | 0.6283(2) | 1.41 | 0.0520(6) | -0.0004(4) | 0.6289(2) | 0.27 |
| Ti(2)' | 0.0533(6) | -0.0001(1) | -0.6286(1) | 6.61 | 0.037(2) | 0.004(2) | 0.3717(2) | 1.41 | 0.0499(6) | 0.0002(4) | 0.3717(2) | 1.77 |
| O(1)   | 0.207(4) | 0.278(5) | 0.4967(8) | 3.22 | 0.322(2) | 0.235(1) | 0.5069(2) | 1.01 | 0.2990(12) | 0.2760(12) | 0.5102(3) | 0.37 |
| O(1)'  | 0.207(4) | -0.278(5) | -0.4967(8) | 3.22 | 0.322(2) | -0.235(1) | 0.4931(2) | 1.01 | 0.3548(11) | -0.2179(11) | 0.4942(3) | 0.59 |
| O(2)   | 0.264(7) | 0.252(9) | 0.2501(7) | 0.91 | 0.265(1) | 0.263(1) | 0.2485(2) | 0.97 | 0.2704(17) | 0.2442(16) | 0.2495(6) | 0.97 |
| O(2)'  | 0.264(7) | -0.252(9) | -0.2501(7) | 0.91 | 0.265(1) | 0.737(1) | 0.7515(2) | 0.97 | 0.2736(16) | 0.7571(16) | 0.7489(6) | 0.65 |
| O(3)   | 0.073(4) | 0.025(6) | 0.5596(8) | 1.27 | 0.086(1) | -0.0640(9) | 0.5594(2) | 0.65 | 0.0913(18) | -0.0705(16) | 0.5605(4) | 0.05 |
| O(3)'  | 0.073(4) | -0.025(6) | -0.5596(8) | 1.27 | 0.086(1) | 0.0640(9) | 0.4406(2) | 0.65 | 0.0918(18) | 0.0587(16) | 0.4424(4) | 0.37 |
| O(4)   | -0.040(4) | 0.074(5) | 0.6815(8) | 1.63 | 0.052(1) | 0.0547(9) | 0.6807(1) | 1.43 | 0.0552(24) | 0.0584(19) | 0.6825(5) | 0.25 |
| O(4)'  | -0.040(4) | -0.074(5) | -0.6815(8) | 1.63 | 0.052(1) | -0.0547(9) | 0.3193(1) | 1.43 | 0.0568(24) | -0.0441(19) | 0.3195(5) | 2.19 |
| O(5)   | 0.294(4) | 0.215(6) | 0.6215(8) | 3.49 | 0.284(2) | 0.247(2) | 0.6109(2) | 0.9 | 0.2904(18) | 0.2800(15) | 0.6121(5) | 0.13 |
| O(5)'  | 0.294(4) | -0.215(6) | -0.6215(8) | 3.49 | 0.284(2) | -0.247(2) | 0.3891(2) | 0.9 | 0.2962(18) | -0.2659(16) | 0.3892(5) | 0.13 |
| O(6)   | 0.159(4) | -0.300(5) | 0.6310(8) | 4.74 | 0.217(2) | -0.299(2) | 0.6244(2) | 0.67 | 0.3677(17) | -0.1959(15) | 0.6244(4) | 0.54 |
| O(6)'  | 0.159(4) | 0.300(5) | -0.6310(8) | 4.74 | 0.217(2) | 0.299(2) | 0.3756(2) | 0.67 | 0.3496(17) | 0.2164(15) | 0.3773(4) | 0.35 |

\* For *B2cb* structure ( )' coordinates are generated by 2-fold rotation about x-axis.

**Table 4: Fractional Coordinates of Structure D and E. For structure E (*B1a1* space group), only initial coordinates (x, y, z) are reported but for structure D (*B2cb* space group), equivalent (x, -y, -z) coordinates are also reported.**

|  | Structure D | | | Structure E | | |
| --- | --- | --- | --- | --- | --- | --- |
| Ion | x | Y | z | x | y | z |
| Bi(1) | 0* | 0.4959 | 0.5662 | 0 | 0.4965 | 0.5664 |
| Bi(1)'** | 0 | 0.5041 | 0.4338 | 0.0009 | 0.4975 | 0.4337 |
| Bi(2) | -0.0002 | 0.4809 | 0.7110 | -0.0067 | 0.4812 | 0.7105 |
| Bi(2)' | -0.0002 | 0.5191 | 0.2890 | -0.0025 | 0.5175 | 0.2884 |
| Ti(1) | 0.0338 | 0* | 0.5* | 0.0353 | -0.0028 | 0.5015 |
| Ti(2) | 0.0448 | -0.0039 | 0.6284 | 0.0446 | -0.0037 | 0.6288 |
| Ti(2)' | 0.0448 | 0.0039 | 0.3716 | 0.0426 | 0 | 0.3715 |
| O(1) | 0.3073 | 0.2480 | 0.5107 | 0.2741 | 0.2862 | 0.5115 |
| O(1)' | 0.3073 | -0.2480 | 0.4893 | 0.3531 | -0.2049 | 0.4935 |
| O(2) | 0.2622 | 0.2464 | 0.2511 | 0.2585 | 0.2462 | 0.2509 |
| O(2)' | 0.2622 | 0.7536 | 0.7489 | 0.2582 | 0.7536 | 0.7490 |
| O(3) | 0.0789 | -0.0774 | 0.5591 | 0.0792 | -0.0768 | 0.5604 |
| O(3)' | 0.0789 | 0.0774 | 0.4409 | 0.0773 | 0.0679 | 0.4428 |
| O(4) | 0.0435 | 0.0574 | 0.6821 | 0.0454 | 0.0582 | 0.6824 |
| O(4)' | 0.0435 | -0.0574 | 0.3179 | 0.0401 | -0.0546 | 0.3181 |
| O(5) | 0.2811 | 0.2801 | 0.6102 | 0.2781 | 0.2842 | 0.6109 |
| O(5)' | 0.2811 | -0.2801 | 0.3898 | 0.2827 | -0.2737 | 0.3900 |
| O(6) | 0.3624 | -0.1941 | 0.6255 | 0.3644 | -0.1922 | 0.6264 |
| O(6)' | 0.3624 | 0.1941 | 0.3745 | 0.3498 | 0.2041 | 0.3762 |

*coordinates kept fixed for comparison of different structure
** For *B2cb* structure ( )' coordinates are generated by 2-fold rotation about x-axis.

**Table 5: Bond lengths (Å) and bond angles (°) in structure D**

**Bismuth Oxide layer**
Bi(2)---O(2) 2.297, 2.399, 2.238
**Perovskite Layer**
Bi(1)---O(3) 2.366, 3.149, 2.343, 3.187
**Central Octahedron**
Ti(1)---O(1) 1.873, 2.036
Ti(1)---O(3) 1.991
O(1)—Ti(1)—O(1) 174.2°
O(3)—Ti(1)—O(3) 165.8°
**Upper Octahedron & Lower Octahedron**
Ti(2)---O(5) 1.971, 2.093
Ti(2)---O(6) 1.918, 2.014
Ti(2)---O(4) 1.784
Ti(2)---O(3) 2.309
O(6)—Ti(2)—O(5) 158.8°, 157.7°
O(3)—Ti(2)—O(4) 175.6°

**Table 6: Bond lengths (Å) and bond angles (°) in structure E**

**Bismuth Oxide layer**
Bi(2)---O(2)            2.238, 2.312
Bi(2)---O(2)'           2.300, 2.416
Bi(2)'---O(2)           2.287, 2.384
Bi(2)'---O(2)'          2.284, 2.234

**Perovskite Layer**
Bi(1)---O(3)            2.355, 3.136, 2.333, 3.181
Bi(1)'---O(3)'          2.379, 3.127, 2.346, 3.163

**Central Octahedron**
Ti(1)---O(1)            1.869, 2.057
Ti(1)---O(1)'           1.883, 2.059
Ti(1)---O(3)            1.991
Ti(1)---O(3)'           1.981
O(1)'—Ti(1)—O(1)        172.4°
O(3)—Ti(1)—O(3)'        166.5°

**Upper Octahedron**
Ti(2)---O(5)            1.961, 2.093
Ti(2)---O(6)            1.915, 2.015
Ti(2)---O(4)            1.794
Ti(2)---O(3)            2.291
O(6)—Ti(2)—O(5)         158.4°, 159.5°
O(3)—Ti(2)—O(4)         175.1°

**Lower Octahedron**
Ti(2)'---O(5)'          1.964, 2.063
Ti(2)'---O(6)'          1.919, 2.005
Ti(2)'---O(4)'          1.780
Ti(2)'---O(3)'          2.380
O(5)'—Ti(2)'—O(6)'      156.8°, 156.4°
O(3)'—Ti(2)'—O(4)'      175.8°